\documentclass[12pt,notitlepage,onecolumn]{revtex4}
\usepackage{graphicx,amssymb,amsmath,subfigure}
\begin{document}

\title{Evolve Networks Towards Better Performance: a Compromise between Mutation and Selection}

\author{Zhen Shao and Hai-jun Zhou\\
Institute of Theoretical Physics,
the Chinese Academy of Sciences,
Beijing 100080, China}

\begin{abstract}
The interaction between natural selection and random mutation is
frequently debated in recent years. Does similar dilemma also exist
in the evolution of real networks such as biological networks? In
this paper, we try to discuss this issue by a simple model system,
in which the topological structure of networks is repeatedly
modified and selected in order to make them have better performance
in dynamical processes. Interestingly, when the networks with
optimal performance deviate from the steady state networks under
pure mutations, we find the evolution behaves as a balance between
mutation and selection. Furthermore, when the timescales of
mutations and dynamical processes are comparable with each other,
the steady state of evolution is mainly determined by mutation. On
the opposite side, when the timescale of mutations is much longer
than that of dynamical processes, selection dominates the evolution
and the steady-state networks turn to have much improved performance
and highly heterogeneous structures. Despite the simplicity of our
model system, this finding could give useful indication to detect
the underlying mechanisms that rein the evolution of real systems.
\end{abstract}

\maketitle

\section{Introduction}
\label{sec:introduction}

The successful application of evolutionary process
\cite{Barabasi-Albert-1999,Watts-Strogatz-1998} to the explanation
of real networks, was one of the major achievements in the research
of complex networks. Meanwhile, the topological structure of complex
networks was frequently found to have determinative effect on the
dynamical processes running on
them\cite{Derrida-Pomeau-1986,Pastor-Satorras-Vespignani-2001,Boccaletti-etal-2002},
and vice versa\cite{Bi-Poo-2001}. Therefore, many dynamical systems
were modeled as adaptive networks, in which feedback from dynamical
processes was extensively coupled into the structural
evolutions\cite{Holme-Newman-2006,Garlaschelli-etal-2007,Gross-Blasius-2008}.
In most of these networks, the compositional elements could actively
change their interaction subjects according to the dynamical states.
But for many biological networks, structural mutation happened
blindly and randomly, with a timescale much separated from that of
dynamical processes\cite{Jacob-1977,Sole-etal-2003}. Besides, the
structural evolution of these systems was mainly driven by
preferential selection, in which systems exhibiting better
performance during the life circle could pass their structural
information to future generations with higher
probability\cite{Darwin-1859}. Thus, an important issue emerged: how
intensively could the pursuit of specific performance, such as
reaching functional states rapidly and
robustly\cite{Siegal-Bergman-2002}, reshape the topological
structures via preferential selection?

In this paper, we investigate this question by designing a simple
model system, in which we evaluate a network by its efficiency of
escaping from disordered states in Local-Majority-Rule (LMR)
dynamical processes\cite{Zhou-Lipowsky-2005} and try to evolve
networks of low efficiency towards networks of high efficiency via
repeated mutations and
selections\cite{Stern-1999,Oikonomou-Cluzel-2004}. Although LMR
dynamics is too simple to represent most of the dynamical processes
running on real systems, it could make the evolution only focus on
optimizing the distribution pattern of edges among vertices, which
greatly reduces the complexity of our problem. In the model system,
we find the steady state of evolution depends strongly on the
timescales of mutations and dynamical processes. When their
timescales are comparable to each other, mutation dominates the
evolution and the steady-state networks have similar structure as
the steady-state networks of pure mutations. On the opposite hand,
when the timescale of mutations is much longer than that of
dynamical processes, selection dominate the evolution and highly
heterogeneous networks with heavily connected hub and much improved
efficiency emerge from the evolution. In the extreme situation,
networks with optimal efficiency, which also deviate significantly
from the steady-state networks of pure mutations in topological
structure. At the end of this paper, we also propose a simple model
on the evolution of particles. Additionally, we illustrate that this
conclusion is still valid even in infinite population limit by a
simple mathematical model. As an extension, this finding calls for a
comprehensive understanding of the evolution of real systems, which
is consistent with the suggestion that the effect of likelihood
should be also incorporated\cite{Whitfield-2007}.

\section{Results}
\label{sec:results}

\begin{figure}
    \begin{center}
    \includegraphics[width=\textwidth]{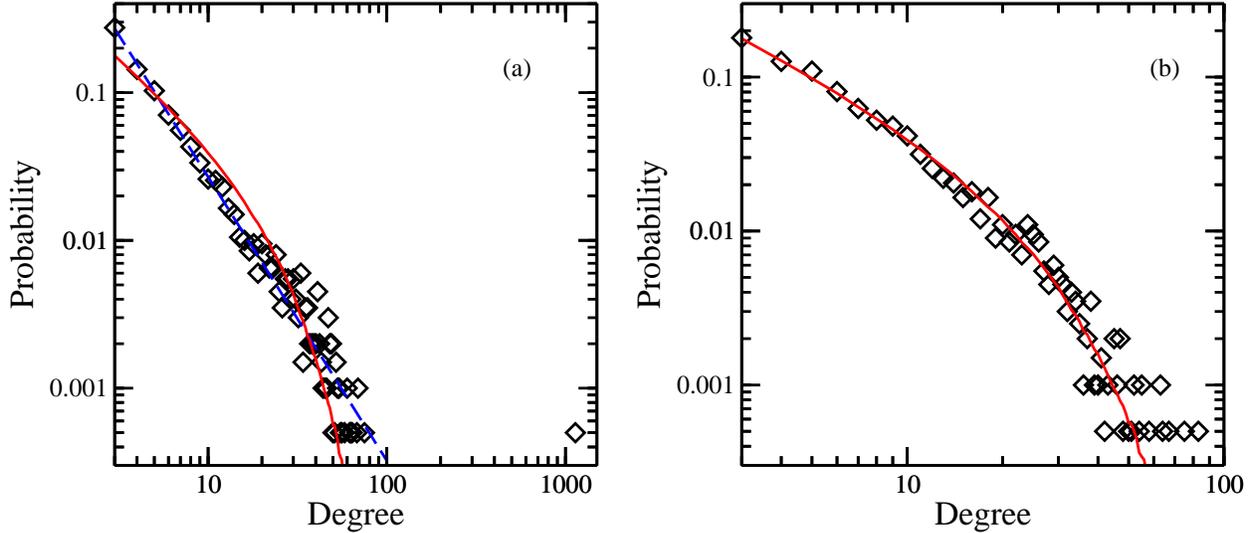}
    \end{center}
    \caption{The steady-state network under dynamics-driven evolution
with different mutation rate $\mu=0.01$ (a) or $\mu=1$ (b) shows
different vertex-degree distributions. Both of the networks are of
size $N=2000$, average connectivity $\langle k \rangle=10$ and
minimal vertex-degree $k_0=3$. The solid lines represent the
corresponding degree distribution for steady-steady networks under
pure mutation (averaged over $200$ samples), while the dashed line
in (a) is the best power-law fit $P(k)\sim k^{-1.92\pm0.02}$ for the
degree distribution of non-global hub vertices.}
\label{fig:degree_distribution}
\end{figure}

In the evolution with low mutation rate ($\mu=0.01$), the networks
evolve to have highly heterogeneous 
structures (Fig.~\ref{fig:degree_distribution}.[a]), in which a
vertex with degree comparable to network size $N$ emerges to act as
the communicating hub.
Meanwhile, the degree distribution of the other vertices shifts to
be power-law-like for small degrees, which is significantly
different from the exponential degree distribution of the
steady-state networks under pure mutation. Besides, these highly
heterogeneous networks exhibit strong degree-degree correlations
indicated by two parameters $R\approx-0.495\pm0.025$, which means
the global hub prefers intensively to interact with vertices having
low degrees, and $r\approx0.067\pm0.0083$, which means the other
vertices prefer to connect with vertices having similar degrees. We
notice that such degree-degree correlations are consistent with the
spin-spin correlations embedded in the strongly disordered spin
configurations of highly heterogeneous networks.
In other words, the dynamics-driven evolution is able to detect the
correlation between different vertices' states and transfer it to
topological clustering, which gives another potential explanation to
the community-rich structures of many biological and social
networks.

\begin{figure}
    \begin{center}
        \includegraphics[width=\textwidth]{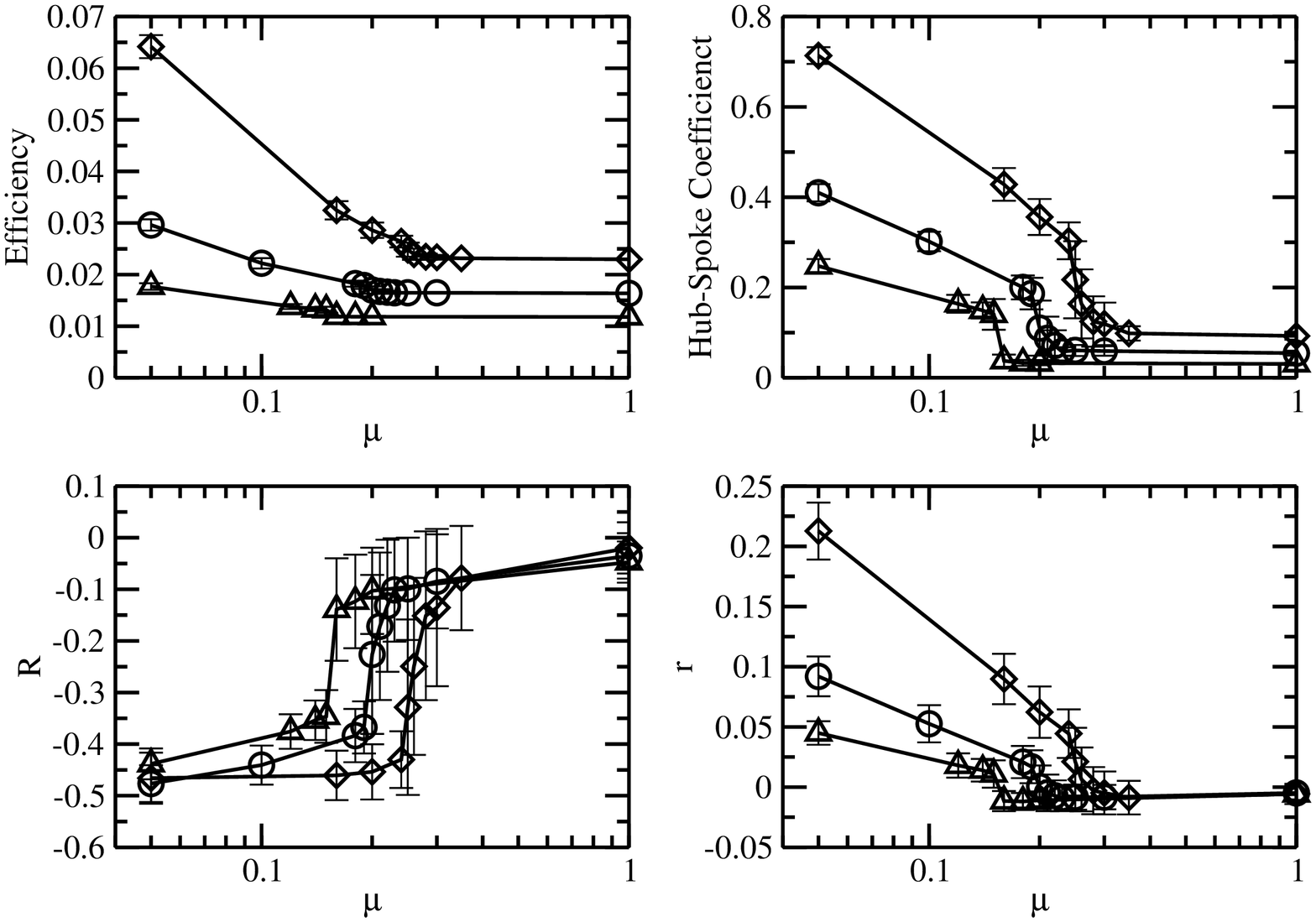}
    \end{center}
        \caption{
        Transition of the steady-state mean efficiency $\mathcal{E}$,
        mean hub-spoke coefficient $C_{hs}$, mean correlation index $R$
        and mean assortative-mixing index (of global hub-removed subnetwork) $r$
        under dynamics-driven evolution versus mutation rate $\mu$. The critical mutation rate
        $\mu_c=0.25$, $0.2$, $0.15$ for network size $N=500$ (diamonds),
        $1000$ (circles), $2000$ (squares), respectively.
        Each point represents a average over $6\times10^4$ generations after the
        steady state of evolution is reached.}
    \label{fig:transition_mu}
\end{figure}

On the opposite side, in the dynamics-driven evolution with high
mutation rate ($\mu=1$) selection fails to achieve further
improvement in both efficiency and structure than mutation-driven
evolution, and the networks remain similar to the steady-state
networks under pure mutations
(Fig.~\ref{fig:degree_distribution}.[b]). From extensive
simulations, we find there exists a critical mutation rate $\mu_c$
which classifies the dynamics-driven evolutions with different
mutation rate into two distinct regimes
(Fig.~\ref{fig:transition_mu}). In the regime $\mu>\mu_c$, mutations
dominate the evolution, and the steady-state networks have similar
structure as the steady-state networks under pure mutation. While in
the regime $\mu<\mu_c$, the steady-state networks appear to have
highly heterogeneous structures, and their improvement in both
efficiency and structure grows monotonically as $\mu$ decreased.
Importantly, the critical mutation rate $\mu_c$ decreases rapidly
with $N$, which indicates a growing difficulty for large networks to
achieve apparent optimization via dynamics-driven evolution.

\begin{figure}
    \begin{center}
        \includegraphics[width=\textwidth]{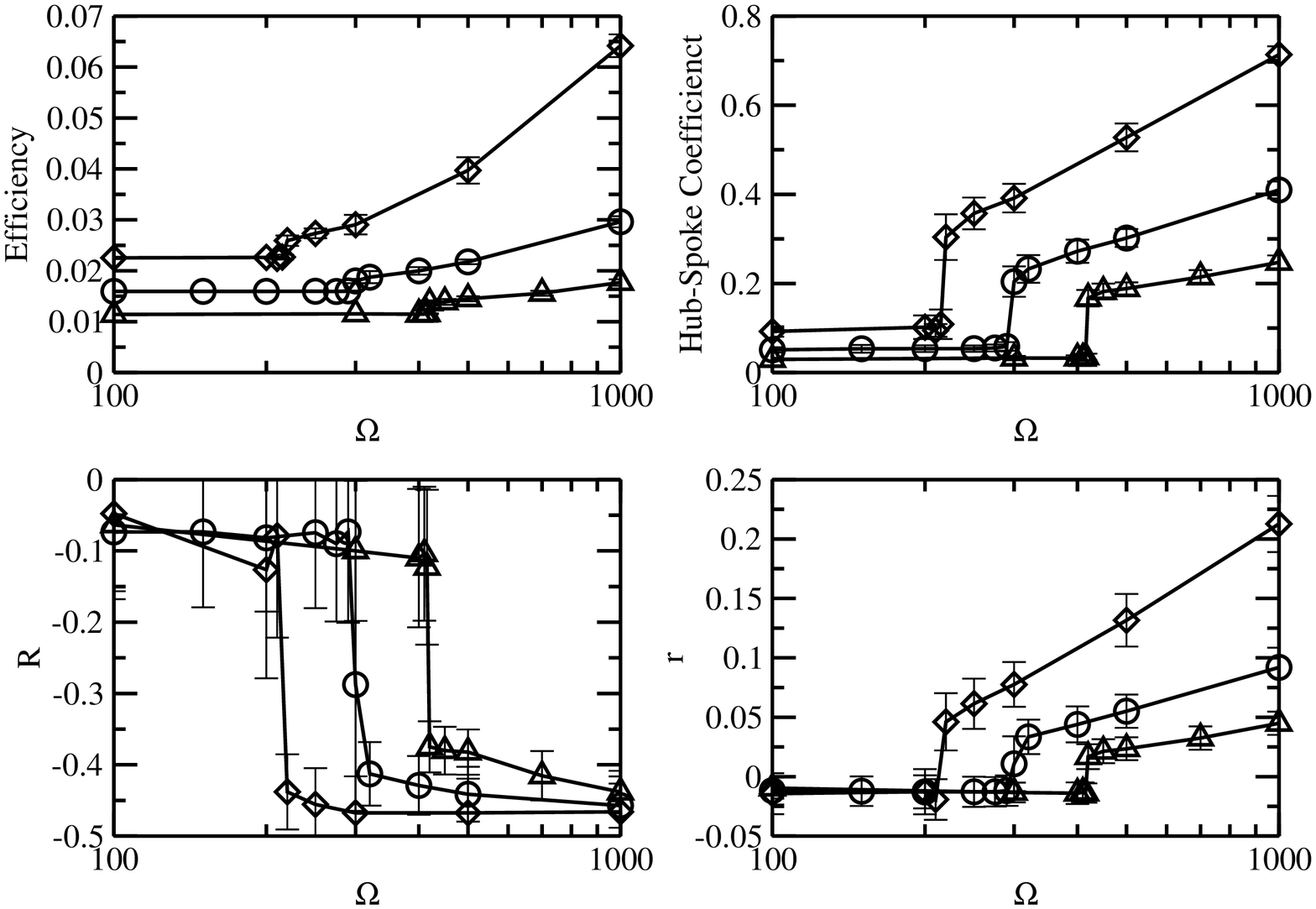}
    \end{center}
        \caption{
        Transition of the steady-state mean efficiency $\mathcal{E}$,
        mean hub-spoke coefficient $C_{hs}$, mean correlation index $R$
        and mean assortative-mixing index (of global hub-removed subnetwork) $r$
        under dynamics-driven evolution versus sampling number $\Omega$. The critical
        sampling number $\Omega_c=220$, $300$, $420$ for network size $N=500$ (diamonds),
        $1000$ (circles), $2000$ (squares), respectively.
        Each point represents a average over $6\times10^4$ generations after the
        steady state of evolution is reached.}
    \label{fig:transition_omega}
\end{figure}

\begin{figure}
    \begin{center}
        \includegraphics[width=\textwidth]{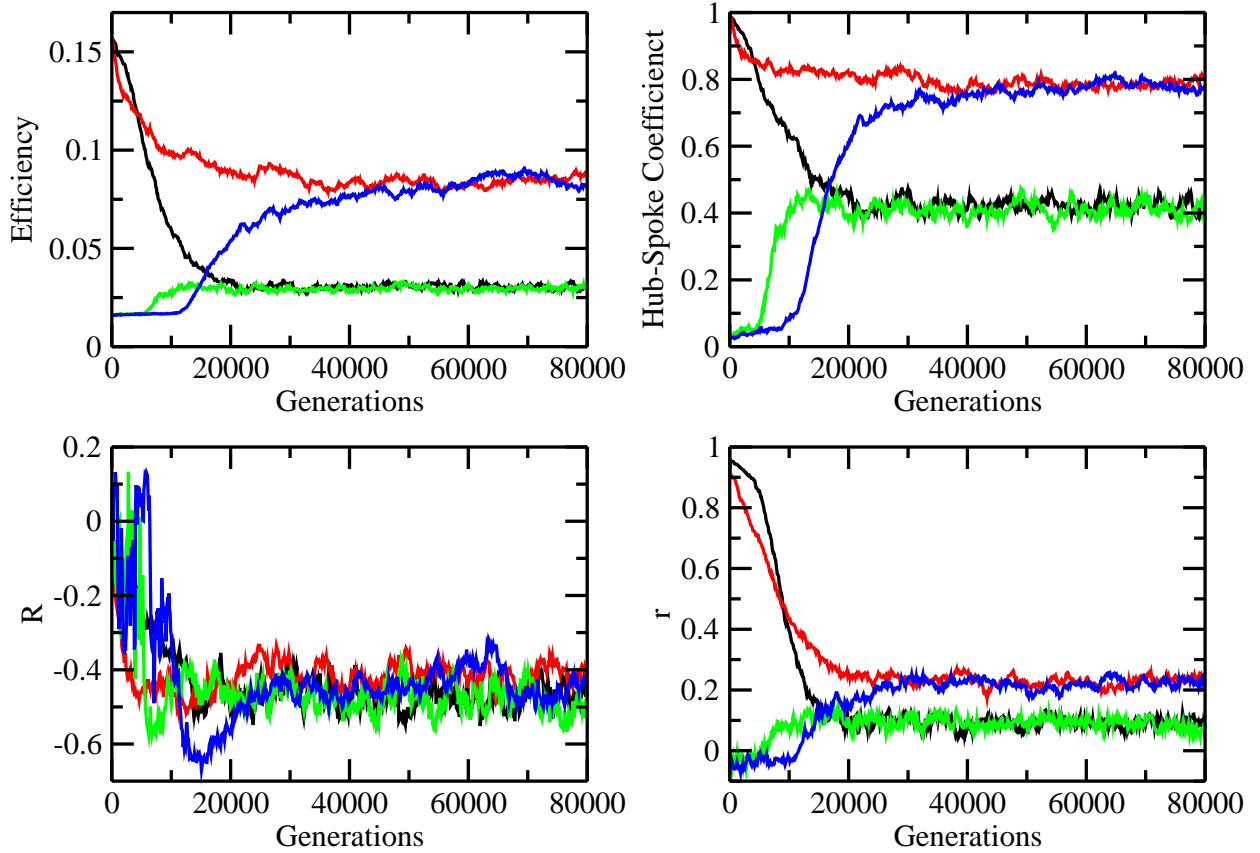}
    \end{center}
        \caption{
        (Color Online) The evolution of mean efficiency $\mathcal{E}$,
        mean hub-spoke coefficient $C_{hs}$, mean correlation index $R$
        and mean assortative-mixing index (of global hub-removed subnetwork) $r$
        as a function of simulation generations. The mutation rate is $\mu=0.05$ (black and
        green curves) or  $\mu=0.01$ (red and blue curves). In the dynamics-driven
        evolution started from hub-spoke
        networks (black and red curves), the population is refreshed mostly by degraded
        networks due to their dominant fraction in the offsprings, and gradually
        converged the steady state of dynamics-driven evolution started from
        random Poissonian networks (green and blue curves).}
    \label{fig:converge-evolution}
\end{figure}

Interestingly, for the dynamics-driven evolutions with different
sampling number $\Omega$, which determines the accuracy of
efficiency sampling, similar transition emerges again
(Fig.~\ref{fig:transition_omega}): only in the regime
$\Omega>\Omega_c$, highly heterogeneous networks emerge and their
improvement grows monotonically with $\Omega$; besides, the critical
sampling number $\Omega_c$ increases rapidly with $N$. Thus, we
conclude that the dynamics-driven evolution is determined by the
balance of two counteracting effects: the promotive effect generated
from preferential selection struggles to push the population towards
networks with high efficiency, while the degradative effect brought
together by random mutation and imprecise sampling of efficiency
drives the population back to the steady-state networks under pure
mutations. Consequently, the steady state of dynamics-driven
evolution should be independent of the original networks, even when
they are optimally designed (Fig.~\ref{fig:converge-evolution}).

To give a mathematical illustration, we consider a population of $M$
particles with two possible energy states $\varepsilon_1=0$ and
$\varepsilon_2=1$. For a particle with energy $\varepsilon_i$ ($i=1$
or $2$), it will be assigned with an random evaluation following
normal distribution
\begin{equation}
\mathcal{N}(\varepsilon_i,\frac{\sigma^2_i}{\Omega})=
\frac{1}{\sigma_i}\sqrt{\frac{\Omega}{2\pi}}e^{-\Omega(x-\varepsilon_i)^2/(2\sigma_i^2)},
\label{eq:normal-distribution}
\end{equation}
here the standard deviation coefficient $\sigma_i$ satisfies
$1\ll\sigma_i^2\ll+\infty$ and $\Omega\in[1,+\infty)$ is the control
parameter. Besides, the offspring particle generated from a particle
with energy $\varepsilon_i$ will appear on the opposite energy level
with probability $p_i$,
otherwise it will have identical energy with its parent. For
mutation-driven evolution, every particle is replaced by an
offspring generated from it at each generation. Easily seen, the
particle composition $r_g$, which is defined to be the fraction of
particles with energy $\varepsilon_2$ at generation $g$, will
finally evolve to a steady state
\begin{equation}
r_s^m=\frac{p_1}{p_1+p_2}.
\label{eq:evolution-equation-1}
\end{equation}
While in evaluation-driven evolution, every particle generates $E$
offsprings ($E$ is positive and finite) at the start of each
generation, then only the $M$ particles with the highest evaluation
of all $(E+1)M$ particles will survive and pass into next
generation. Therefore, in the infinite population limit
($M\rightarrow\infty$), the steady-state particle composition
$r_s^e$ will satisfy
\begin{equation}
\left\{\begin{array}{ll}r_s^e=[r_s^e+E(1-r_s^e)p_1+Er_s^e(1-p_2)]
\int_{\phi_s}^{+\infty}\mathcal{N}(\varepsilon_2,\frac{\sigma_2^2}{\Omega})dx \\
1-r_s^e=[1-r_s^e+E(1-r_s^e)(1-p_1)+Er_s^ep_2]
\int_{\phi_s}^{+\infty}\mathcal{N}(\varepsilon_1,\frac{\sigma_1^2}{\Omega})dx\end{array}\right.,
\label{eq:evolution-equation-2}
\end{equation}
here $\phi_s$ is the steady-state threshold that only particles with
evaluation higher than it could survive. When
$\Omega\rightarrow+\infty$, which means the stochastic fluctuation
in particle evaluation will decrease to vanish, we find that
$\int_{\phi_s}^{+\infty}\mathcal{N}(\varepsilon_1,\frac{\sigma_1^2}{\Omega})dx\rightarrow
\int_{\varepsilon_2}^{+\infty}\mathcal{N}(\varepsilon_1,0)dx=0$ and
$r_s^e\rightarrow1$. Oppositely, when $\Omega$ is sufficient small
so that
\begin{equation}
\int_{\phi_s}^{+\infty}\mathcal{N}(\varepsilon_1,\frac{\sigma_1^2}{\Omega})dx\approx
\int_{\phi_s}^{+\infty}\mathcal{N}(\varepsilon_2,\frac{\sigma_2^2}{\Omega})dx,
\label{eq:survive-prob-equal}
\end{equation}
which implies particles with different energy will survive with
approximately equal probability, the steady-state composition will
have $r_s^e\approx\frac{p_1}{p_1+p_2}=r_s^m$. On the other hand, in
the extreme condition $p_1\rightarrow0$ and $p_2\rightarrow1$, which
corresponds to the dynamics-driven evolution with high mutation
rate, it could be derived from equation
(\ref{eq:evolution-equation-1}) and equation
(\ref{eq:evolution-equation-2}) that $r_s^e\rightarrow0$ and
$r_s^m\rightarrow0$. In other words, the steady-state population of
evaluation-driven evolutions has similar composition with that of
mutation-driven evolution. However, in the extreme condition
$p_1\approx p_2\ll1$, which corresponds to the dynamics-driven
evolution with small mutation rate, we find that $r_s^e\approx1$ and
$r_s^m\approx\frac{1}{2}$. Obviously, this compositional difference
indicates an apparent optimization achieved by
evaluation-driven evolution. 
Thus, if we take the dynamics-driven evolution as particles jumping
among a series of structural heterogeneity levels, it could be
easily derived from above analysis that the optimal structure could
be approached only when the timescale of mutations is much longer
than that of dynamical processes\cite{Shao-Zhou-2008}.

\begin{figure}
    \begin{center}
        \includegraphics[width=\textwidth]{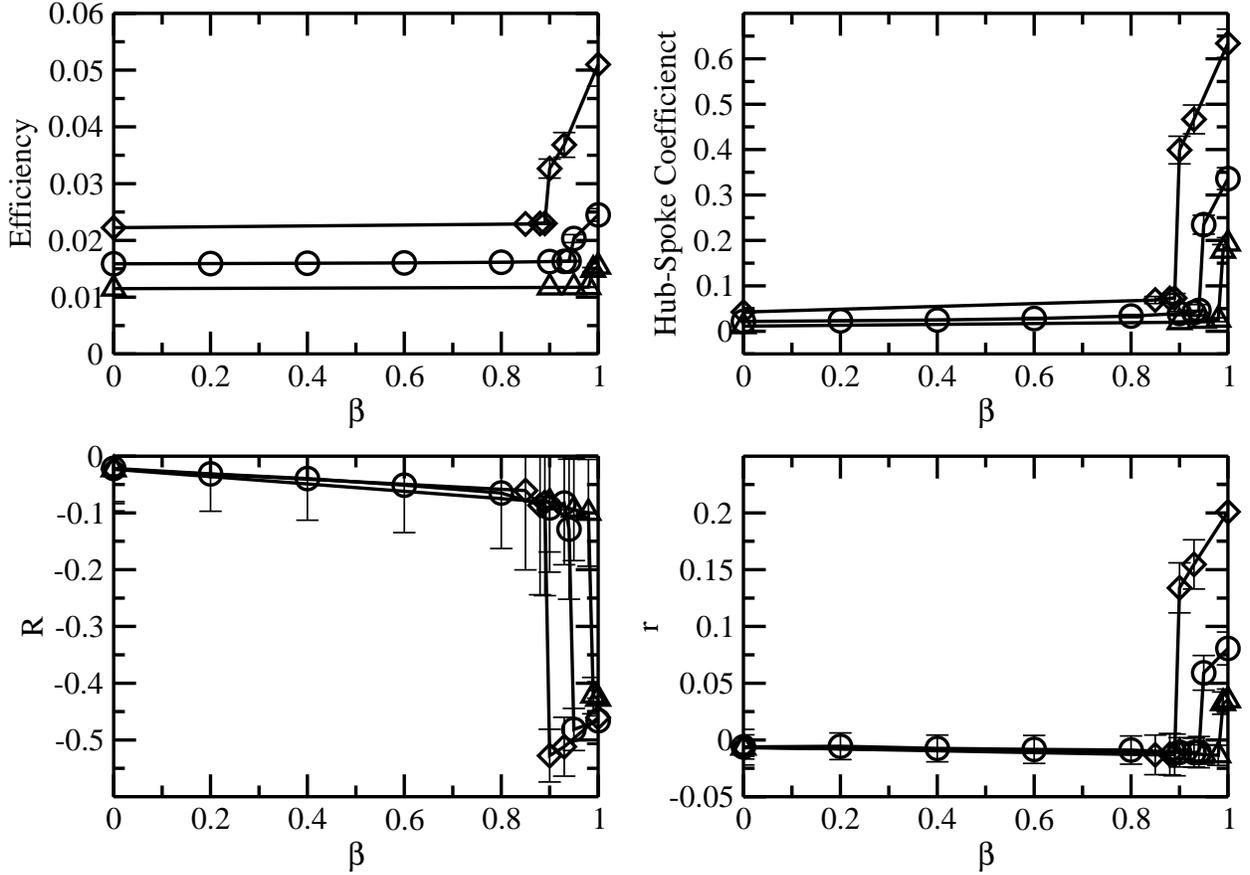}
    \end{center}
        \caption{
        Transition of the steady-state mean efficiency $\mathcal{E}$,
        mean hub-spoke coefficient $C_{hs}$, mean correlation index $R$
        and mean assortative-mixing index (of global hub-removed subnetwork) $r$
        under dynamics-driven evolution versus the strength of preferential rewiring $\beta$.
        The critical strength $\beta_c=0.9$, $0.95$, $0.99$ for network size $N=500$ (diamonds),
        $1000$ (circles), $2000$ (squares), respectively.
        Each point represents a average over $6\times10^4$
        generations after the steady state of evolution is reached.}
    \label{fig:transition_beta}
\end{figure}

Finally we replace the local rewiring scheme by preferential
rewiring, and show there also exists a critical strength $\beta_c$
(Fig.~\ref{fig:transition_beta}) to classify dynamics-driven
evolutions with different strength of preferential rewiring into
mutation-dominating regime ($\beta<\beta_c$) or selection-dominating
regime ($\beta>\beta_c$). Interestingly, the critical strength
$\beta_c$ is found to increase monotonically with $N$ as well, which
reemphasizes the growing difficulty to achieve apparent optimization
in large networks and supports that our finding is independent of
the way how mutation occurs.

\section{Discussion}
\label{sec:discussion}

It has been debated whether the evolution in biological world should
be viewed as some hill-climbing process\cite{Wright-1982}, or is
mainly determined by the "neutral mutations"\cite{Kimura-1983} like
a random walk. But in our simple model system, we find when the
optimal networks deviate from the steady state networks under
mutations, a flexible combination of these two viewpoints is
required. Interestingly, similar deviations are frequently detected
in biological and social systems, such as the abundance of certain
network motifs\cite{Milo-etal-2002} or nucleotide
types\cite{Hallin-Ussery-2004}. Therefore, dynamics-driven evolution
could be a potential way to explain these structural
heterogeneities. However, the evolution of real systems is far more
complicated than our simple model system. Meanwhile, the evaluation
of them is usually determined by a combination of multiple
components such as efficiency, sensitivity, robustness and
cost\cite{Mathias-Gopal-2001}. Therefore, more detailed models are
still needed to give a comprehensive understanding to the evolution
running in real world.

\section{Methods}
\label{sec:methods}

\subsection{Local-Majority-Rule dynamics}

For a undirected network of size $N$ and average connectivity $c$,
we attach to each vertex $i$ a binary spin variable $\sigma_i=\pm
1$. At each time step $t$ of the LMR dynamics, all the vertices
update their states synchronously according to
\begin{equation}
\sigma_i(t)={\rm sign}[\sum\limits_{j\in\partial i}\sigma_{j}(t-1)],
\label{eq:LMR-rule}
\end{equation}
here $\partial i$ means the vertices which are connected with vertex
$i$ directly. As described in \cite{Shao-Zhou-2008}, we start a LMR
dynamical process from a strongly disordered configuration
${\vec{\sigma}}(0) \equiv \{\sigma_1(0), \sigma_2(0),\ldots,
\sigma_N(0) \}$, which is randomly generated with the constraints
$\sum\limits_{i=1}^N \sigma_{i} = \sum\limits_{i=1}^N k_i \sigma_{i}
= 0$ ($k_i$ is the degree of vertex $i$). Typically, the LMR
dynamics will drive the network to stabilize at a consensus state,
in which all the vertices have the same spin state. It has been
shown in \cite{Zhou-Lipowsky-2005} that in LMR dynamical processes,
scale-free networks with degree distribution $P(k)\sim k^{-\gamma}$
and $\gamma\rightarrow2$ can reach consensus more rapidly than
Poissonian networks of similar size and average connectivity. In
other words, highly heterogeneous networks are more efficient in
communicating internal states than homogenous networks.

\subsection{Definition of efficiency}

For a given network ${\cal G}$, we start a total number of $\Omega$
different LMR dynamical processes, and run them for $T=1$ time step
to reach the corresponding configuration $\vec{\sigma}^{\alpha}(1)$.
It has been shown in \cite{Zhou-Lipowsky-2005}
that the characteristic relaxation time of a network in the LMR
dynamics is determined by the the escaping velocity of the network's
spin configuration from the strongly disordered region. In this
paper, we define the efficiency ${\cal G}$ of network ${\cal G}$ as
the average change of vertex-state in $\vec{\sigma}(1)$, which is
calculated according to
\begin{equation}
\mathcal{E}({\cal G})= \frac{1}{\Omega}
\sum\limits_{\alpha=1}^{\Omega} \Biggl| \frac{1}{N}
\sum\limits_{i=1}^{N} \sigma^\alpha_i(1) \Biggr| \ .
\label{eq:efficiency-definition}
\end{equation}
Easily seen, the efficiency sampled in this way
is only a random estimation of the true efficiency $\mathcal{E}^*({\cal
G})=\lim\limits_{\Omega\rightarrow\infty}\mathcal{E}({\cal G})$.
According to the Central Limit Theorem, the distribution of
$\mathcal{E}({\cal G})$ will be approximately
Gaussian-distribution-like with the average value equal to
$\mathcal{E}^*({\cal G})$, while the mean square variance will be
proportional to $1/\Omega$ (see the Supplementary Information).
That's to say, the efficiency of a network will be evaluated more
precisely with large $\Omega$.

\subsection{Evolutionary algorithm}

The dynamics-driven evolution starts from a population of $P$
networks, which are uniformly sampled from the ensemble of random
Poissonian networks of given size $N$ and average connectivity
$\langle k\rangle$. At the beginning of each generation, every
network generates $E$ exact copies, which expands the population to
a size (E + 1)P. Then every new network undergoes random mutations
following local rewiring\cite{Baiesi-Manna-2003}:
\begin{enumerate}
\item[(1)] For each vertex, mutation occurs with probability
$\mu$ (mutation rate);

\item[(2)] if mutation is accepted by vertex $i$, a rewiring of edge
$(i,j)\rightarrow(i,l)$ is proposed, in which vertex $j$ is randomly
selected from the nearest neighbors of vertex $i$ and vertex $l$ is
then randomly selected from the nearest neighbors of vertex $j$
except vertex $i$;

\item[(3)] this proposal will be rejected if and only if edge $(i,l)$ already exists or
the degree of vertex $j$ is less than a minimal value $k_0$ after
cutting edge $(i, j)$.
\end{enumerate}
Then the efficiency of each network is sampled independently, and
the population shrink to its original size with only $P$ networks
with the highest efficiency survive and pass into next generation.
As a reference to the dynamics-driven evolution, we also derive the
steady-state networks under pure mutations by proposing a
mutation-driven evolution. In the evolution, all the networks keep
to be replaced by a new network generated from it following the same
rules of mutation. For the local rewiring scheme, the degree
distribution of the steady-state networks under pure mutations is
shown to decay exponentially\cite{Baiesi-Manna-2003}. If not
specified by the context, the parameters default to adopt value
$\Omega=1000$, $P=25$, $N=1000$, $\langle k\rangle=10$ , $E=3$ and
$k_0=5$.

In this work we also involved preferential rewiring
scheme\cite{Ohkubo-Yasuda-2005}: the only change from above
description is that vertex $l$ ($l\neq i$ and $l\neq j$) in the
rewiring proposal is chosen randomly with a probability $\Pi_l
\propto (k_l+1)^\beta$, here $k_l$ is vertex $l$'s degree and
$\beta$ is the control parameter which determines the strength of
preferential rewiring. Obviously when $\beta=0$, the preferential
rewiring scheme will degrade to random rewiring, under which the
steady-state networks has been proved to be
Poissonian\cite{Ohkubo-Yasuda-2005}. On the opposite side, the
maximal value of $\beta$ used in this paper is $1$, under which the
steady-state networks will have exponential degree
distributions\cite{Ohkubo-Yasuda-2005}.

\subsection{Identification of global hub and Definition of hub-spoke coefficient}

For generality, we identify the vertex with the highest degree in a
network as the global hub for both heterogeneous and homogenous
networks. (If there are multiple vertices with the highest degree,
we only identify the one with the smallest index as the only global
hub.) We also define the hub-spoke coefficient $C_{hs}$ of a network
to be $C_{hs}=\frac{k_{hub}}{N-1}$, in which $k_{hub}$ is the global
hub's degree and $N$ is network size. Obviously, when $C_{hs}$
equals to $1$, the network will look completely hub-and-spoke like.

\subsection{Quantification of degree-degree correlation in networks}

To give an explicit quantification of the degree-degree correlations
in a network, we define two quantities: correlation index $R$
measures the degree-degree correlation on links connecting the
global hub and the other vertices, while assortative-mixing index
$r$ measures the degree-degree correlations on links connecting the
other vertices. We define $R$ to be the normalized ratio of the mean
degree of nearest neighbors of the global hub $\langle k_{nn}^{hub}
\rangle$ to the averaged value $\langle k_{nn}^{hub} \rangle_{ran}$
of this mean degree over an ensemble of randomly shuffled
networks\cite{Milo-etal-2002}, and calculate it according to
\begin{equation}
R = \left\{ \begin{array}{ll} ( \langle k_{nn}^{hub} \rangle -
\langle k_{nn}^{hub} \rangle_{ran})/(\langle k_{nn}^{hub}
\rangle_{max} - \langle k_{nn}^{hub}\rangle_{ran} ),
\;\mathrm{if}\; \langle k_{nn}^{hub} \rangle >  \langle k_{nn}^{hub}\rangle_{ran} \\
( \langle k_{nn}^{hub} \rangle - \langle k_{nn}^{hub}
\rangle_{ran})/(\langle k_{nn}^{hub} \rangle_{min} - \langle
k_{nn}^{hub}\rangle_{ran} ), \;\mathrm{if}\; \langle k_{nn}^{hub}
\rangle < \langle k_{nn}^{hub}\rangle_{ran} \\
0,\;\mathrm{if}\; \langle k_{nn}^{hub} \rangle = \langle
k_{nn}^{hub}\rangle_{ran}
\end{array} \right.,
\label{eq:R-definition}
\end{equation}
here $\langle k_{nn}^{hub} \rangle_{max}$ and $\langle k_{nn}^{hub}
\rangle_{min}$ are the maximal and minimal value of $\langle
k_{nn}^{hub} \rangle$ in the ensemble of randomly shuffled networks,
respectively. Obviously, $R\in[-1,1]$. When $R<0$, the global hub
prefers to interact with vertices with low degrees compared with
its behavior in the randomly shuffled networks. Otherwise, when
$R>0$, vertices with high degrees are preferred. On the other hand,
we define $r$ to be the assortative-mixing index of the globe
hub-removed subnetwork following \cite{Newman-2002}
\begin{equation}
r = \frac{ M'\sum\limits_{i=1}^{M'}j_ik_i -
\left[\sum\limits_{i=1}^{M'} \frac{j_i+k_i}{2}\right]^2 } {
M'\sum\limits_{i=1}^{M'}\frac{j_i^2+k_i^2}{2} -
\left[\sum\limits_{i=1}^{M'} \frac{j_i+k_i}{2}\right]^2 },
\label{eq:rr-definition}
\end{equation}
here $M'$ is the number of edges in the subnetwork.
When $r>0$, these vertices prefer to be assortatively connected
with vertices having similar degrees; Otherwise, when
$r<0$, the subnetwork is disassortatively connected.

To make rapid calculation of $R$, here we give a theoretical
prediction of $\langle k_{nn}^{hub}\rangle_{ran}$ by
\begin{equation}
\langle k_{nn}^{hub}\rangle_{ran} =
\frac{1}{k_{hub}}\sum\limits_{k=1}^{k_{hub}}k(N_k-\delta_{k,k_{hub}})P^*(k),
\label{eq:R-prediction}
\end{equation}
in which $N_k$ is the number of vertices with degree $k$, $\delta$
is the Kronecker symbol and $P^*(k)$ is the probability that a
vertex with degree $k$ is connected with the global hub in a
randomly shuffled network. Now we approximate $P^*(k)$ by
\begin{eqnarray}
P^*(k) \approx
\sum\limits_{i=1}^{k_{hub}}\left\{P^*_i(k)\prod\limits_{j=1}^{i-1}[1-P^*_j(k)]\right\},
\end{eqnarray}
in which
\begin{eqnarray}
P^*_i(k)=\frac{k}{2M-k_{hub}-\sum\limits_{j=1}^{i-1}\left\{\sum\limits_{k'=1}^{k_{hub}}
[k'(N_{k'}-\delta_{k',k_{hub}})P^*_j(k')\prod\limits_{m=1}^{j-1}(1-P^*_m(k'))]\right\}}.
\end{eqnarray}
In numerical simulations, it's shown to be a good approximation for
the networks appeared in this paper (see the Supplementary
Information).

\subsection{Generation of hub-spoke networks}

A hub-spoke network of size $N$, average
connectivity $c$ and minimal degree $k_0$ is generated through:

\begin{enumerate}
\item[(1)] first generate a random Poissonian network of size $N-1$ and average connectivity
$c'=c-\frac{N-1}{N}$;

\item[(2)] then evolve the Poissonian network under mutation-driven evolution
with minimal degree $k'_0=k_0-1$ until the steady state is reached;

\item[(3)] then assortatively shuffle the mutated network
following \cite{Xulvi-Brunet-Sokolov-2004} with $p=1$;

\item[(4)] finally add a global hub to the shuffled network
and connected it with all the other $N-1$ vertices.
\end{enumerate}
From numerical simulations, we find such hub-spoke networks have
much higher efficiency than the Poisson networks of the same size
and average connectivity (Fig.~\ref{fig:converge-evolution}).

\section*{Acknowledge}

We thank Ming Li, Kang Li and Jie Zhou for helpful discussions, and
Zhong-Can Ou-Yang for support. We benefited from the KITPC 2008
program "Collective Dynamics in Information Systems".

\clearpage

\end{document}


\title{Supplementary Information}

\author{Zhen Shao and Hai-jun Zhou\\
Institute of Theoretical Physics,
the Chinese Academy of Sciences,
Beijing 100080, China}

\maketitle

\section{Local-Majority-Rule dynamics}

A strongly disordered configuration could be constructed as follows:
one first divide the vertices into two groups ($G_+$ and $G_-$) of
equal size randomly and assign spin $+1$ to vertices in group $G_+$
and assign spin $-1$ to vertices in group $G_-$; then as long as
$\sum\limits_{i=1}^N k_i \sigma_{i} \neq 0$, two vertices (one from
$G_+$ and the other from $G_-$) are chosen randomly and their states
are exchanged if and only if this exchange does not cause an
increase in the value of $|\sum\limits_{i=1}^N k_i \sigma_{i}|$. In
such a way, a vertex (either chosen randomly or reached by following
a randomly chosen edge) will be found to be in the spin $+1$ state
with probability $1/2$.

Easily seen, there exist clear spin-spin correlations in the
strongly disordered configurations of highly heterogeneous networks.
For a network, we define the probability distribution $q(k)$ as the
probability that a vertex with degree $k$ is found to have the same
spin state as the global hub in the strongly disordered
configurations. Then we find for highly heterogenous networks,
$q(k)$ is clearly higher than $1/2$ when $k$ is small and is clearly
lower than $1/2$ when $k$ is large
(Fig.~\ref{fig:spin_correlation}). That's to say, the degree-degree
correlations in the highly heterogenous networks reached by dynamics
driven evolution are consistent with the spin-spin correlations
embedded in the strongly disordered configurations of these
networks.

However, these spin-spin correlation aren't the main causation of
the emergence of highly heterogenous networks in the dynamics-driven
evolution. To prove it, we modify the model system by using
simplified LMR dynamics. In the simplified LMR dynamical processes,
the initial spin configurations are disordered configurations which
only need to satisfy the constraint $\sum\limits_{i=1}^N
\sigma_{i}=0$. In this paper, a disordered configuration is
constructed by dividing the vertices into two groups of equal size
randomly and assigning opposite spin states to the vertices in
different groups. From numerical simulations, we find the
dynamics-driven evolution with low mutation rate still could reach
highly heterogeneous networks, but the degree-degree correlations in
them are much weaker (Fig.~\ref{fig:simplified_dynamics}).

\begin{figure}
    \begin{center}
    \includegraphics[width=0.5\linewidth]{Spin_Correlation_N_2000.eps}
    \end{center}
        \caption{
        The probability distribution $q(k)$ of a network of which the degree
        distribution is the same as the one shown by Fig.~1.(a)
        (red circles) of main text or the one shown by Fig.~1.(b)
        (black diamonds) of main text.}
    \label{fig:spin_correlation}
\end{figure}

\begin{figure}
    \begin{center}
    \includegraphics[width=0.8\linewidth]{Evolution_Simplified_Dynamics.eps}
    \end{center}
        \caption{
        Evolution of mean efficiency $\mathcal{E}$, mean hub-spoke coefficient $C_{hs}$,
        mean hub preference $R$ and mean assortative-mixing index (of hub-excluded subnetwork)
        $r$ driven by simplified LMR dynamics (black curves). The evolution
        driven by standard LMR dynamics (red curves) is also plotted for comparison.}
    \label{fig:simplified_dynamics}
\end{figure}

\section{Definition of efficiency}

\begin{figure}
    \begin{center}
        \includegraphics[width=\textwidth]{Efficiency_Variance_2.eps}
    \end{center}
        \caption{
        (a). The distribution of $10000$ independent
        samplings of a random Poissonian network's efficiency.
        The efficiency is sampled with $\Omega=1000$,
        and the network is of size $N=1000$ and average connectivity $\langle k\rangle=10$.
        (b). Mean square variance of the efficiency
        distribution of the same network versus $\Omega$. These results is consistent
        with the prediction from the Central Limit Theorem.}
    \label{fig:efficiency_sampling}
\end{figure}

\begin{figure}
    \begin{center}
    \includegraphics[width=\linewidth]{Efficiency_Network_Type_2.eps}
    \end{center}
        \caption{
        (a). The efficiency of different kinds of networks. Networks
        of type $1$ are scale-free networks of exponent $\gamma=2.003$.
        Networks of type $2$ are scale-free networks of exponent
        $\gamma=2.24$. Networks of type $3$ are scale-free networks of exponent
        $\gamma=2.498$. Networks of type $4$ are scale-free networks of exponent
        $\gamma=2.81$. Networks of type $5$ are scale-free networks of exponent
        $\gamma=3.234$. Networks of type $5$ are random Poissonian networks.
        All the networks are of size $N=1000$ and average
        connectivity $\langle k\rangle=10$.
        (b). The positive correlation between the efficiency and the
        maximum degree of a scale-free network. The scale-free networks
        are of exponent $\gamma=2.003$ (black diamonds),
        $\gamma=2.498$ (red circles), $\gamma=2.81$ (blue triangles). In both
        of the graphs the efficiency is sampled with $\Omega=1000$.}
    \label{fig:efficiency_versus_network_type}
\end{figure}

To check the validity of our definition of efficiency in this work,
we first measure the efficiency of different kinds of networks
including scale-free networks and Poisson networks. As shown in
Fig.~\ref{fig:efficiency_versus_network_type},  we find when
scale-free networks, which could be characterized by their power-law
degree distributions $P(k)\sim k^{-\gamma}$, have much higher
efficiency than random Poisson networks of equal size and average
connectivity. Additionally, this advantage in efficiency grows as
the exponent $\gamma$ decreases approach $2$. These results are
consistent with the finding in
\cite{Zhou-Lipowsky-2005,Zhou-Lipowsky-2007} that scale-free
networks of $\gamma\rightarrow 2$ have much lower characteristic
relax time in LMR dynamics than random Poisson networks of equal
size and average connectivity. On the other hand, we also find the
characteristic relax time of the steady-state networks under
dynamics-driven evolutions decreases as the their efficiency
grows(Fig.~\ref{fig:transition_median_decay_time}). Thus, we could
conclude the efficiency defined in this paper is sufficient to
represent the velocity of a network to reach consensus in LMR
dynamical processes. Besides, we find the efficiency of the scale-free
networks is positively correlated with the maximum degree of the network.
This finding is consistent with the emergence of global hub
led by dynamics-driven evolution. In other words, our model system reemphasizes
the conclusion reached by many works that heavily connected hubs
play important roles in the evolution of many dynamical
processes\cite{Morgan-Soltesz-2007}.

Consequently, one may argue that the efficiency of a network in LMR
dynamics could be directly evaluated by the characteristic relax
time. However, because the stochastic deviation in the sampling of
characteristic relax time is much larger than that in efficiency
sampling, the mutation-dominating regime expands dramatically and
apparent improvement is significantly more difficult to be achieved
by dynamics-driven evolution.

\begin{figure}
    \begin{center}
        \includegraphics[width=0.5\textwidth]{Transition_Median_Decay_Time_no_color.eps}
    \end{center}
        \caption{
        The average median decay time\cite{Zhou-Lipowsky-2005} of the steady-state networks
        reached by the dynamics-driven evolutions with different mutation rate $\mu$
        (a), different sampling number $\Omega$ (b)
        or different strength of preferential rewiring (c).
        The networks are of size $N=500$ (diamonds),
        $1000$ (circles), $2000$ (squares), and the median decay time $\tau_D$
        of a network is calculated as follows: first run $\Omega=1000$
        independent LMR dynamical processes on the network; then $\tau_D$ is the time
        at which more than half of the $\Omega$ processes reach
        $|\sum\limits_{i=1}^N k_i\sigma_i(\tau_D)|\geq \sum\limits_{i=1}^N k_i/2$.}
    \label{fig:transition_median_decay_time}
\end{figure}

\section{Quantification of degree-degree correlations in networks}

\begin{figure}
    \begin{center}
    \includegraphics[width=0.6\linewidth]{Definition_R_check_scatter.eps}
    \end{center}
        \caption{
        Examination of the rapid calculation of correlation index $R$.
        The quantity $R$ of each network is calculate according to
        equation (7) of main text, in which $\langle k_{nn}^{hub}\rangle_{ran}$
        is predicted by equation (9) of main text, while the
        quantity $R'$ of each network is calculate according to equation (\ref{eq:R-prime-definition}).
        The networks are uniformly selected from the dynamics-driven evolution
        with mutation rate $\mu=0.2$ (red cycles), $0.05$ (cyan squares),
        $0.01$ (green diamonds), $0.005$ (blue triangles). The black line is the
        complementary curve $R'=R$.}
    \label{fig:R_check}
\end{figure}

To examine the theoretical approximation used in the rapid calculation of correlation
index $R$, we define quantity $R'$ of a network as
\begin{equation}
R' = \left\{ \begin{array}{ll} ( \langle k_{nn}^{hub} \rangle -
\langle k_{nn}^{hub} \rangle^*_{ran})/(\langle k_{nn}^{hub}
\rangle_{max} - \langle k_{nn}^{hub}\rangle^*_{ran} ),
\;\mathrm{if}\; \langle k_{nn}^{hub} \rangle >  \langle k_{nn}^{hub}\rangle^*_{ran} \\
( \langle k_{nn}^{hub} \rangle - \langle k_{nn}^{hub}
\rangle^*_{ran})/(\langle k_{nn}^{hub} \rangle_{min} - \langle
k_{nn}^{hub}\rangle^*_{ran} ), \;\mathrm{if}\; \langle k_{nn}^{hub}
\rangle < \langle k_{nn}^{hub}\rangle^*_{ran} \\
0,\;\mathrm{if}\; \langle k_{nn}^{hub} \rangle = \langle
k_{nn}^{hub}\rangle^*_{ran}
\end{array} \right.,
\label{eq:R-prime-definition}
\end{equation}
, in which $\langle k_{nn}^{hub}\rangle^*_{ran}$ is the average
$\langle k_{nn}^{hub}\rangle$ value of $400$ randomly shuffled
networks of given network and the other quantities are the same as
they are in equation (7) of main text. From Fig.~\ref{fig:R_check}
we find these two quantities of the networks appeared in the
dynamics-driven evolution meet well. So we could conclude that the
approximation used in the rapid calculation of $R$ is valid for both
homogenous networks and highly heterogeneous networks.

\section{Evolutionary algorithm}

\begin{figure}
    \begin{center}
    \includegraphics[width=0.5\linewidth]{Evolution_Expanding_Coefficient.eps}
    \end{center}
        \caption{
        (Color Online) The evolution of mean efficiency $\mathcal{E}$,
        mean hub-spoke coefficient $C_{hs}$, mean correlation index $R$
        and mean assortative-mixing index (of global hub-removed subnetwork) $r$
        with different expanding coefficient $E=1$ (black curves), $3$
        (red curves), $7$ (blue curves).}
    \label{fig:expanding_coefficient}
\end{figure}

\begin{figure}
    \begin{center}
        \includegraphics[width=\textwidth]{Transition_Population_Size_no_color.eps}
    \end{center}
        \caption{
        Transition of the steady-state mean efficiency $\mathcal{E}$,
        mean hub-spoke coefficient $C_{hs}$, mean correlation index $R$
        and mean assortative-mixing index (of global hub-removed subnetwork) $r$
        under dynamics-driven evolution versus population size $P$. The critical
        population size $\P_c=9$, $12$, $15$ for network size $N=500$ (diamonds),
        $1000$ (circles), $2000$ (squares), respectively.
        Each point represents a average over $6\times10^4$ generations after the
        steady state of evolution is reached.}
    \label{fig:transition_population_size}
\end{figure}

We find the expanding coefficient $E$ has no effect on the steady state
of dynamics-driven evolution, but only determines the velocity to
reach steady state: the evolution with large expanding coefficient could
reach steady state more rapidly that that with low expanding coefficient
(Fig.~\ref{fig:expanding_coefficient}). On the other hand, we find the
dynamics-driven evolution with different population size could also be
classify into two regimes by a critical population size $P_c$:
mutation-dominating regime $P<P_c$ and selection-dominating regime
$P>P_c$ (Fig.~\ref{fig:transition_population_size}).
Consistent with the findings introduced in the main part of this paper,
the critical population size $P_c$ also increases with network size $N$.

\section{Evolution process}

\begin{figure}
    \begin{center}
        \includegraphics[width=\textwidth]{Evolution_mutation_rate.eps}
    \end{center}
        \caption{
        The evolution process of mean efficiency $\mathcal{E}$,
        mean hub-spoke coefficient $C_{hs}$, mean correlation index $R$
        and mean assortative-mixing index (of global hub-removed subnetwork) $r$
        with different mutation rate $\mu=1$ (black curves), $0.2$ (red curves),
        $0.05$ (green curves), $0.01$ (blue curves), $0.005$ (magenta curves).}
    \label{fig:evolution_mutation_rate}
\end{figure}

\begin{figure}
    \begin{center}
        \includegraphics[width=\textwidth]{Evolution_test_trials.eps}
    \end{center}
        \caption{
        The evolution process of mean efficiency $\mathcal{E}$,
        mean hub-spoke coefficient $C_{hs}$, mean correlation index $R$
        and mean assortative-mixing index (of global hub-removed subnetwork) $r$
        with different sampling number $\Omega=100$ (black curves), $290$ (red curves),
        $300$ (green curves), $1000$ (blue curves), $5000$ (magenta curves).}
    \label{fig:evolution_test_trials}
\end{figure}

From the detailed evolution process
(Fig.~\ref{fig:evolution_mutation_rate} and
Fig.~\ref{fig:evolution_test_trials}), we could find that only in
the dynamics-driven evolution in the selection-dominating regime,
highly heterogeneous networks are able to dominating the population
continuously. Besides, the velocity to reach steady state in the
evolution processes is find to positively related with mutation rate
and sampling number.

\clearpage